\documentclass[onecolumn]{IEEEtran}
\usepackage{amsmath,amsfonts}
\usepackage{algorithmic}
\usepackage{algorithm}
\usepackage{array}
\usepackage[caption=false,font=normalsize,labelfont=sf,textfont=sf]{subfig}
\usepackage{textcomp}
\usepackage{stfloats}
\usepackage{url}
\usepackage{verbatim}
\usepackage{graphicx}
\usepackage{cite}
\usepackage{caption}

\begin{document}

\title{A Hybrid Gauss–Markov–LSTM Mobility Model for Indoor OWC\\
}

\author{\IEEEauthorblockN{ Walter Zibusiso Ncube, Ahmad Adnan Qidan, Taisir El-Gorashi and Jaafar M. H. Elmirghani} \\
\IEEEauthorblockA{\textit{Department of Engineering, Faculty of Natural, Mathematical and Engineering Sciences} \\
\textit{King’s College London, United Kingdom}\\
(walter.ncube, ahmad.qidan, taisir.elgorashi, jaafar.elmirghani)@kcl.ac.uk}
}

\maketitle

\begingroup
\renewcommand\thefootnote{}
\footnotetext{979-8-3195-4420-9/26/\$31.00~\copyright~2026 IEEE}
\endgroup

\begin{abstract}
 Optical wireless communication (OWC) has emerged as a promising candidate for future high-capacity indoor wireless networks, driven by its large unregulated spectrum, high spatial reuse, and ability to support multi-gigabit data rates. However, OWC systems are highly sensitive to user mobility, as link performance depends strongly on the spatial alignment between transmitter and receiver. Accurate modelling of user position and device orientation is therefore essential for reliable channel estimation and system evaluation. To that effect, this paper proposes a hybrid Gauss--Markov and long short-term memory (GM--LSTM) mobility model for indoor OWC environments. The Gauss--Markov component captures the temporal correlation of user motion, while the LSTM learns residual behaviour to model non-linear movement patterns and orientation dynamics. The proposed model jointly predicts user position and device orientation, enabling improved representation of mobility in OWC channels. Performance is evaluated using prediction accuracy and per-user data rate evolution. Results show that the proposed hybrid GM--LSTM model outperforms conventional Random Waypoint and Gauss--Markov models, providing more accurate mobility prediction and more stable communication performance in dynamic indoor environments.
\end{abstract}


\section{Introduction}

\IEEEPARstart{O}ptical wireless communication (OWC) has emerged as a viable solution for high-capacity indoor wireless connectivity due to its large unregulated spectrum and ability to support highly directional transmission \cite{TB, Walt2, WaltAhrar}. In particular, laser-based systems such as vertical-cavity surface-emitting lasers (VCSELs) enable narrow optical beams and high spatial reuse, making them suitable for next-generation indoor networks \cite{VCSEL}. However, unlike RF systems, where propagation is relatively robust to movement, OWC links are strongly affected by user mobility \cite{RWP}. Variations in user position and, more critically, device orientation can lead to rapid fluctuations in channel gain, resulting in unstable data rates and potential link outages. Therefore, accurate modelling of user mobility, including both spatial movement and orientation dynamics, is essential for reliable performance evaluation and system design.

The Random Waypoint (RWP) model is a widely adopted mobility model in wireless network analysis, in which users move by randomly selecting a destination within the coverage area and travelling towards it with a randomly chosen speed. Upon reaching the destination, a pause time may be introduced before a new destination and speed are selected \cite{RWP_GM}. This process results in a memoryless
mobility pattern, where each movement decision is independent of past motion. This model has been extensively used due to its simplicity and analytical tractability. In \cite{mobb2}, a mobility-aware optical RWP model integrated with transfer learning was developed to enable efficient handover and load balancing in hybrid LiFi--Wi-Fi networks, achieving enhanced throughput, reduced delay, and improved energy efficiency. Similarly, in \cite{RWP}, an orientation-based Random Waypoint (ORWP) model was proposed to incorporate device orientation into mobility analysis, enabling more realistic performance evaluation in LiFi and high-frequency wireless systems. Despite these extensions, RWP fundamentally assumes independent and memoryless movement between consecutive steps \cite{RWP_GM}, which does not accurately reflect realistic indoor user behaviour. In contrast, the Gauss--Markov (GM) mobility model introduces temporal correlation into the movement process by updating user velocity and direction based on their previous values, resulting in smoother and more continuous trajectories. This correlation allows the GM model to better capture the inertia and gradual changes observed in human motion \cite{GM_channel}. In \cite{mobb1}, a GM model was developed for wireless networks to enhance mobility prediction, minimise data loss, and improve network performance in dynamic UAV environments. However, the GM model remains limited in capturing complex human behaviour, particularly abrupt direction changes and device orientation variations. More recently, data-driven approaches based on artificial neural networks (ANNs), have been explored for mobility prediction \cite{ANN_mob,ANN_mob2}. In \cite{ANN_mob1}, an indoor mobility prediction model for wireless networks was developed using a Markov Chain enhanced with Q-learning, enabling accurate user trajectory prediction and facilitating seamless predictive handovers in dense 5G environments. Nevertheless, despite their ability to learn complex non-linear patterns from data, learning-based models often lack robustness to unseen scenarios and require large training datasets for reliable performance.

To address these limitations, this paper proposes a hybrid Gauss--Markov and LSTM (GM--LSTM) mobility model for indoor OWC systems. The GM component provides a structured representation of motion dynamics with temporal correlation, while the LSTM learns residual behaviour that captures deviations from the stochastic model, including orientation changes and non-linear movement patterns. The proposed model jointly predicts user position and device orientation, enabling more accurate representation of mobility in OWC environments. Performance is evaluated using prediction accuracy and per-user data rate evolution. The proposed approach is compared against conventional RWP and GM models, demonstrating improved mobility representation and more reliable communication performance under dynamic user movement.

\section{System Model}

\begin{figure}[ht]
\centering 
\includegraphics[width=0.7\columnwidth]{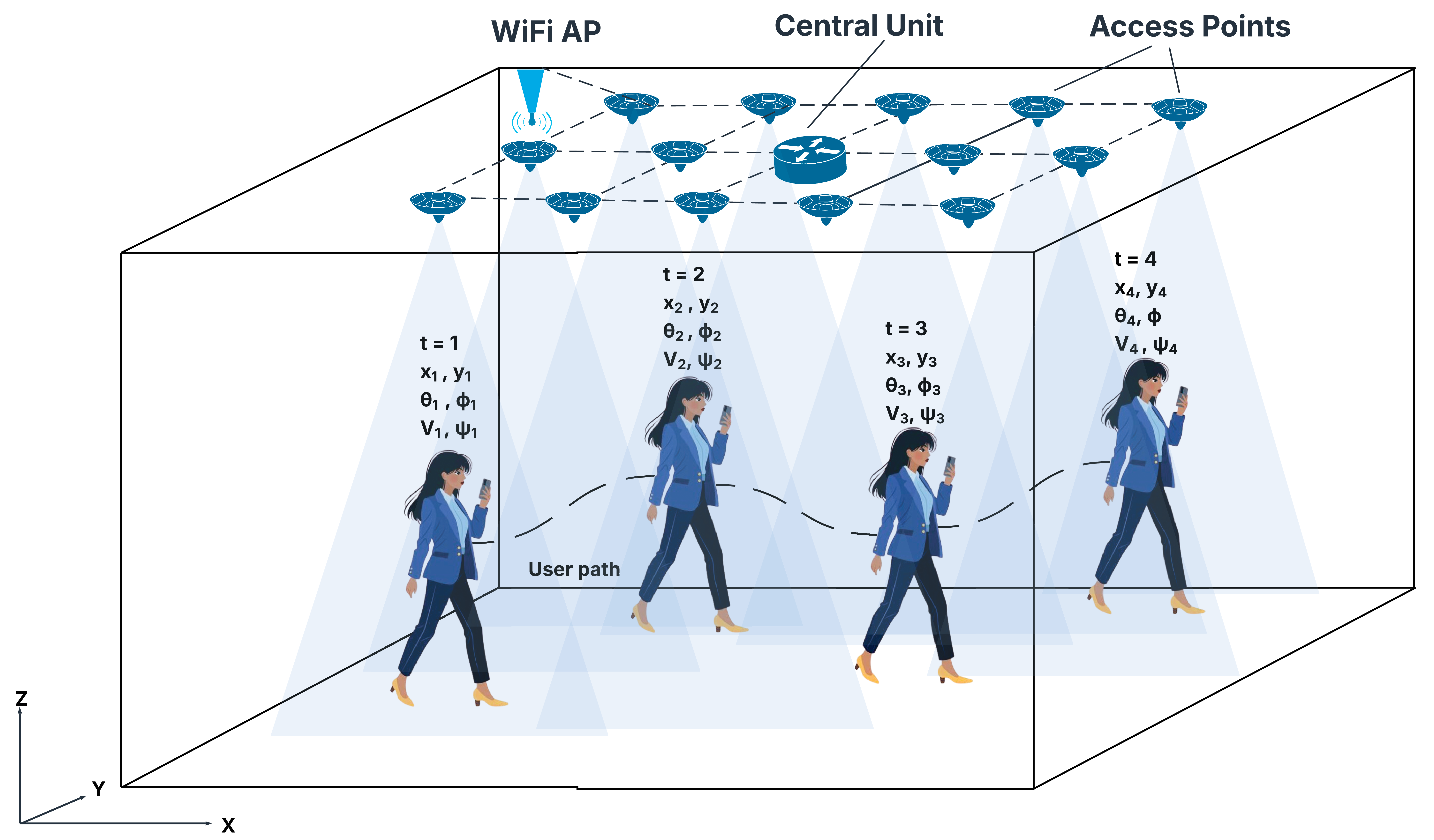}
\caption{System Model} 
\label{fig:System Model} 
\end{figure}

Consider an indoor OWC system as shown in Fig.~\ref{fig:System Model}, with a room of dimensions \(L \times W \times H\) m\(^3\), where a set of ceiling-mounted optical access points (APs), denoted by \(\mathcal{A}=\{1,\ldots,A\}\), provides wireless coverage to a set of mobile users, denoted by \(\mathcal{U}=\{1,\ldots,U\}\). The APs employ VCSEL-based transmitters, and users are equipped with optical receivers on handheld devices. A central control unit coordinates the network operation. In addition, a Wi-Fi AP is employed to support uplink transmission, while the OWC links are primarily used for downlink data delivery. Due to the directional nature of transmission, link quality depends strongly on the relative geometry between APs and users. The state of user \(u\) at time \(t\) is represented by
\begin{equation}
\mathbf{s}_u(t)=[x_u(t),y_u(t),v_u(t),\psi_u(t),\theta_u(t),\phi_u(t)]^T,
\end{equation}
where \(x_u(t)\) and \(y_u(t)\) denote the user position in the horizontal plane, \(v_u(t)\) is the user speed, \(\psi_u(t)\) is the movement direction, and \(\theta_u(t)\) and \(\phi_u(t)\) denote the receiver orientation angles. Since OWC links are highly sensitive to both user location and receiver orientation, accurate modelling of \(\mathbf{s}_u(t)\) is essential for reliable channel representation.

For a given user--AP pair, the optical channel gain \(H_{u,a}(t)\) depends on the propagation distance and the alignment between the transmitter beam and the receiver. Hence, mobility-induced variations in user position and device orientation directly affect the received optical power and achievable data rate. Assuming an intensity modulation and direct detection (IM/DD) system, the achievable data rate for user \(u\) served by AP \(a\) at time \(t\) is given by
\begin{equation}
R_{u,a}(t)=B\log_2\left(1+\frac{(P_tR_{PD} H_{u,a}(t))^2}{\sigma^2 + I_{u,a}(t)}\right),
\end{equation}
where \(B\) is the system bandwidth, $R_{PD}$ is the responsivity, \(P_t\) is the transmitted optical power, \(\sigma^2\) is the receiver noise current variance, and \(I_{u,a}(t)\) represents the aggregate interference current variance from neighbouring APs. Let \(\hat{\mathbf{s}}_u(t+1)\) denote the predicted state of user \(u\) at time \(t+1\). This predicted state is used to determine user-AP associations and compute the estimated channel gain \(\hat{H}_{u,a}(t+1)\), from which the corresponding achievable data rate, $\hat{R}_{u,a}(t+1)$, can be determined.

\section{Hybrid Gauss--Markov and LSTM model }

A hybrid GM--LSTM mobility model is proposed to capture both the temporal correlation of user motion and behaviour-driven mobility dynamics in indoor OWC systems. The model jointly predicts user position and device orientation by combining a stochastic mobility process with a data-driven learning component. The predicted mobility state of user \(u\) at time \(t+1\) is: 
\begin{equation}
\hat{\mathbf{s}}_u(t+1)=f_{\mathrm{GM}}\big(\mathbf{s}_u(t)\big)+f_{\mathrm{LSTM}}\big(\mathbf{s}_u(t-k:t)\big),
\end{equation}
where \(\mathbf{s}_u(t)\) denotes the current mobility state, \(f_{\mathrm{GM}}(\cdot)\) represents the GM prediction, and \(f_{\mathrm{LSTM}}(\cdot)\) is an LSTM-based model that learns the residual mobility behaviour from a sequence of past states of length \(k\). The GM component provides a structured baseline for mobility evolution by introducing temporal correlation into the motion process. It captures the inertia of user movement, ensuring smooth transitions in position, velocity, and direction. However, due to its linear and stochastic nature, it cannot accurately represent complex human behaviour such as abrupt trajectory changes and variations in device orientation.

To address this limitation, an LSTM network is employed to model the residual dynamics not captured by the GM process. The LSTM takes as input a sequence of past mobility states \(\mathbf{s}_u(t-k:t)\), including both positional and orientation information, and outputs a correction term that refines the baseline prediction. The network consists of stacked LSTM layers followed by a fully connected output layer that maps the hidden representation to the state residual. This structure enables the model to capture temporal dependencies and non-linear mobility patterns over multiple time steps. The LSTM is trained offline using supervised learning to learn the residual mobility dynamics, defined as the difference between the true mobility state and the GM prediction at the next time step. By learning this residual, the network focuses on behaviour-driven deviations that are not captured by the stochastic model, rather than attempting to model the full motion dynamics. Training data is generated using synthetic mobility trajectories that represent realistic indoor user behaviour, including variations in speed, direction, and device orientation as in \cite{mobility_abrupt, mobility_abrupt1,mobility_abrupt2, mobility_synth1, mobility_data}. The generated trajectories represent the ground-truth mobility states, corresponding to the actual user movement in terms of position and device orientation. These ground-truth states serve as the target values for prediction. For each time step, the GM model produces a baseline estimate of the next mobility state based on the current state. The LSTM is then trained to learn the residual, defined as the difference between the ground-truth mobility state and the GM prediction for that particular time step. In this way, the GM model captures the structured motion dynamics, while the LSTM learns the additional correction required to account for behaviour-driven deviations. This approach enables the generation of diverse and controlled training data without requiring extensive real-world measurements, while still capturing key mobility characteristics relevant to OWC systems. During inference, the GM model first produces a baseline estimate of the next mobility state. The trained LSTM then predicts the residual correction based on recent mobility history, and the final state estimate is obtained by combining the two components. The control unit then uses the final user state to determine user-AP associations and channel gain estimations.

\section{Performance Evaluation}

This section evaluates the effectiveness of the proposed mobility framework in dynamic indoor OWC environments. The proposed model is compared with the RWP and conventional GM mobility models to assess prediction accuracy and its impact on channel estimation and communication performance under realistic user mobility conditions.

Consider a $5 \times 5 \times 3$ m$^3$ indoor environment with ceiling-mounted optical APs employing VCSEL transmitters. A line-of-sight optical channel is assumed, reflecting the dominant propagation path in OWC systems. The system periodically updates the current state of the user and predicts the next state over a time interval known as the prediction horizon. This prediction horizon is set to 100 ms. User speeds in this system are varied from 0.2 m/s to 1.6 m/s to capture a wider range of realistic mobility conditions. This range accounts for slow movement in highly constrained scenarios as well as faster walking in less restricted environments \cite{user_speed}. Device orientations change dynamically to emulate realistic handheld usage \cite{8790655}. Results are averaged over multiple independent mobility realisations to ensure statistical reliability. To ensure an unbiased and reproducible evaluation, realistic ground-truth user trajectories are generated using a hybrid GM-based simulator with behavioural variations. The GM component captures the temporal correlation of human movement, while bounded variations introduce irregular variations in speed, direction, and device orientation to emulate realistic indoor mobility. These trajectories are used to compute the true optical channel gains, against which the prediction accuracy of the RWP, GM, and proposed hybrid GM--LSTM models is evaluated.

\begin{figure}[ht]
\centering
\subfloat[Channel prediction RMSE vs prediction horizon]{
    \includegraphics[width=0.45\columnwidth]{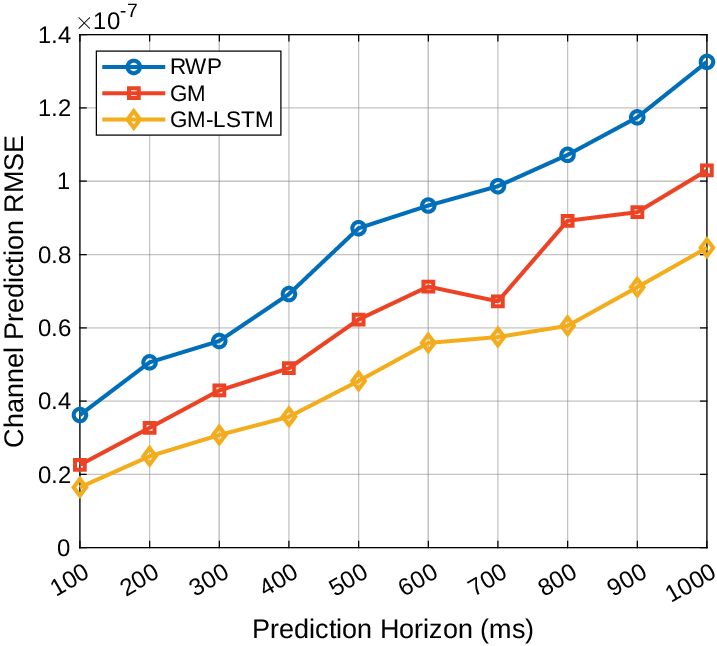}
    \label{fig:CHvsPH}
}
\hfill
\subfloat[Channel prediction RMSE vs user speed]{
    \includegraphics[width=0.45\columnwidth]{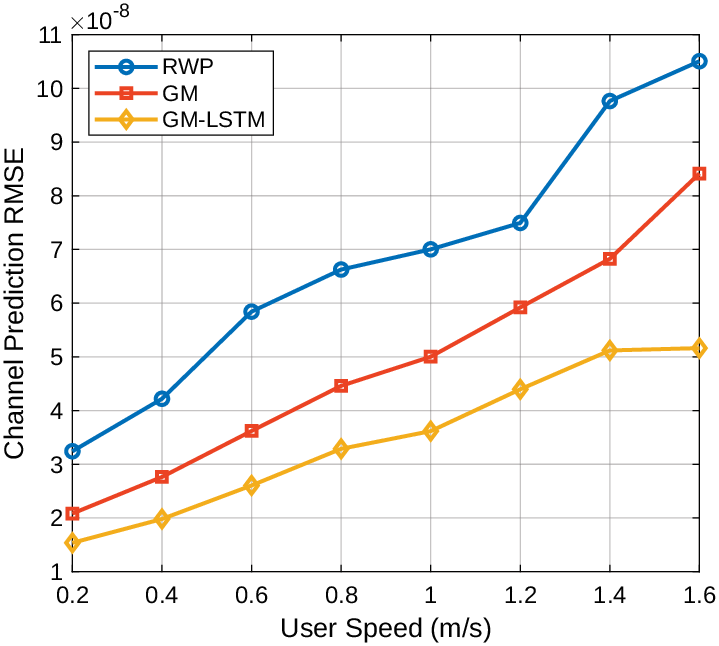}
    \label{fig:CHvsUS}
}
\caption{Channel prediction RMSE versus prediction horizon and user speed}
\label{fig:combined}
\end{figure}

Performance is assessed using three metrics: channel prediction root mean square error (RMSE) versus prediction horizon, channel prediction RMSE versus user speed, and user data rate versus time. The first metric evaluates how prediction accuracy degrades as the prediction horizon increases, while the second examines the robustness of the mobility models under varying mobility conditions. Both of these are shown in Fig. \ref{fig:combined}. The third metric assesses the impact of mobility prediction on communication performance.

Fig.~\ref{fig:CHvsPH} shows that the channel prediction RMSE increases with prediction horizon for all models due to growing uncertainty in user movement over time. The RWP model exhibits the highest error owing to its memoryless nature, which prevents it from capturing temporal dependencies in mobility. In contrast, the GM model achieves improved accuracy by exploiting temporal correlation, resulting in smoother trajectory estimation. The hybrid GM--LSTM framework consistently achieves the lowest RMSE, demonstrating its ability to capture both linear motion dynamics and nonlinear behavioural variations, thereby providing more accurate and reliable mobility prediction.

Fig.~\ref{fig:CHvsUS} illustrates the impact of user speed on channel prediction accuracy. As user speed increases from 0.2 m/s to 1.6 m/s, the prediction error rises for all schemes due to more rapid channel variations and reduced predictability of user movement. Nevertheless, the hybrid GM--LSTM model maintains superior robustness and consistently outperforms the benchmark approaches. This performance gain is attributed to its ability to learn complex mobility patterns and compensate for deviations from the stochastic GM model, enabling more accurate channel estimation under dynamic conditions.

\begin{figure}[ht]
\centering 
\includegraphics[width=0.5\columnwidth]{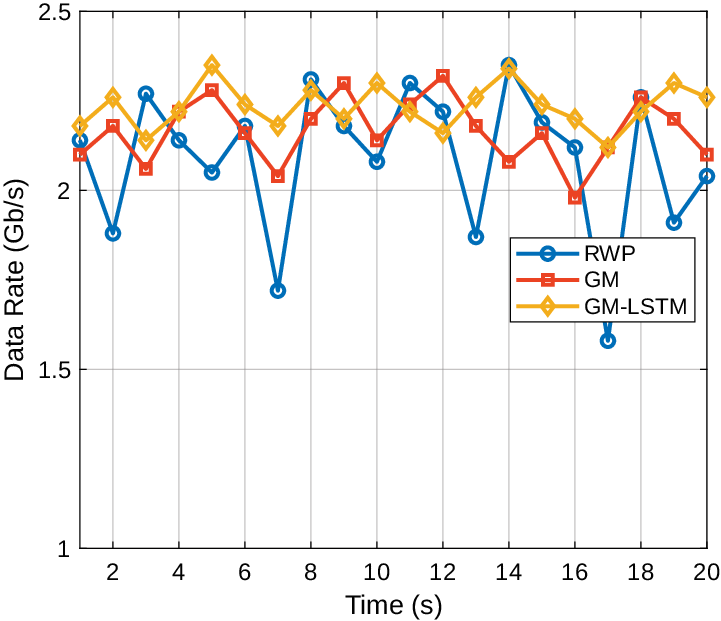}
\caption{Data rate vs Time (Speed = 1.0 m/s)} 
\label{fig:DR Performance} 
\end{figure}

Fig.~\ref{fig:DR Performance} presents the achievable data rate of a user over time, moving at a constant speed of 1.0 m/s. The x-axis represents time in seconds, showing how the data rate evolves as the user moves within the indoor environment under different mobility models. The RWP model exhibits significant fluctuations, with frequent spikes and deep drops, due to its memoryless mobility pattern that leads to inaccurate channel prediction and unstable link adaptation. The GM model provides smoother variations by exploiting temporal correlation, resulting in fewer severe degradations. In contrast, the hybrid GM--LSTM framework delivers the most stable data rate, remaining consistently closer to the maximum achievable throughput while minimising fluctuations. This improvement is attributed to its ability to accurately predict both linear motion and non-linear behavioural changes, leading to more reliable channel estimation and improved link adaptation. Although all schemes occasionally achieve similar peak data rates, the proposed approach significantly reduces performance variability, which is critical for maintaining reliable communication and quality of service in dynamic OWC environments.

\section{Conclusions and Future Work}

This paper presented a hybrid GM--LSTM mobility model for indoor OWC systems. The proposed framework combines a stochastic Gauss--Markov mobility model with a data-driven residual learning approach to jointly capture temporal correlation and behaviour-driven mobility dynamics, including device orientation. By improving the accuracy of mobility modelling, the proposed method enables more reliable estimation of optical channel variations. Performance evaluation demonstrated that the hybrid GM--LSTM model achieves lower channel prediction error and provides more stable data rate performance compared to conventional RWP and GM models. These results highlight the effectiveness of integrating model-based and learning-based approaches for mobility prediction, and their potential to enhance reliability and quality of service in dynamic indoor OWC environments. Future work will focus on extending the proposed framework to more complex and realistic scenarios, including the impact of user blockage, dynamic environmental changes, and multipath propagation. In addition, the integration of the proposed mobility model with adaptive resource allocation and mobility-aware handover strategies will be investigated to further improve overall system performance.

\bibliographystyle{IEEEtran}
\bibliography{References}

\end{document}